\newcommand{\la}{\langle}
\newcommand{\ra}{\rangle}

\newcommand{\beq}{\begin{eqnarray}}
\newcommand{\eeq}{\end{eqnarray}}

\newcommand{\eps}{\epsilon}

\renewcommand{\theequation}{\thesection.\arabic{equation}}

\newcommand{\do}{^{\circ }}

\newcommand{\pipi}{$\pi$-$\pi$\, }

\newcommand\con{\langle \bar {q} q \rangle }

\newcommand{\btem}{\bibitem}

\newcommand{\TK}{T.\ Kunihiro}
\newcommand{\MPL}{Phys.\ Lett.\ {\bf B}}
\newcommand{\MPTP}{Prog.\ Theor.\ Phys.}
\newcommand{\MPR}{Phys.\ Rev.}

\newcommand{\THK}{T. Hatsuda and T. Kunihiro}
\newcommand{\KH}{T. Kunihiro and T. Hatsuda}
\documentstyle[seceq,epsf,wrapft,twoside]{ptptex}
\notypesetlogo  
\title{  Chiral Transition in QCD and  Scalar
Correlations
}
\author{%
Teiji {\sc Kunihiro}
}
\inst{%
Yukawa Institute for Theoretical Physics, Kyoto University,\\
Sakyoku, Kyoto 606-8502, Japan
}
\abst{%
We review recent developments
in exploring possible precursory phenomena of
partial  restoration of  chiral symmetry in nuclear medium by examining
 the spectral function in the scalar channel. 
We emphasize that the wave function renormalization of the 
pion in the medium plays an essential role to induce the
decrease of the pion decay constant as the order parameter
of chiral transition. We also show that the
$\sigma$ meson pole in the complex energy plane moves toward 
the origin as chiral symmetry is restored in hot and/or
dense nuclear matter.
}

\begin{document}
\maketitle

\setcounter{tocdepth}{4}

\section{Introduction}

Hadrons are
 low-lying elementary excitations
 on top of the non-perturbative QCD vacuum, although
QCD Lagrangian itself is written in terms of 
quark- and gluon-fields of which hadrons are composed.
Furthermore,  symmetries possessed by the QCD Lagrangian
are not manifest in the low-energy regime.
These complication are owing to the fact that 
the true QCD vacuum is
realized through the phase 
transitions, i.e.,
 the confinement-deconfinement and the chiral transitions.
Thus  one recognizes that hadron spectroscopy
based on QCD must be a study of the nature of QCD vacuum.

In the present report, focusing on the
chiral transition in hot and/or dense hadronic matter,
I will discuss some characteristic changes in the scalar 
 correlations associated with
the (partial) restoration of chiral symmetry
 in the hadronic medium.
The present report is more or less based on
 \citen{hk94,ptp95,protvino,yitp00,paris,chiral02}.

\setcounter{equation}{0}
\renewcommand{\theequation}{\arabic{section}.\arabic{equation}}

\section{Chiral symmetry restoration as  a phase 
transition of QCD vacuum}

The true QCD vacuum is realized through 
the chiral transition, which is clearly demonstrated 
by lattice simulations\cite{karsch}.

A heuristic argument based on 
a Hellman-Feynman theorem  also tells us that the chiral condensate 
$\con$ decreases at finite density $\rho_{_B}$ as well as at 
finite $T$\cite{DL90}.
For   the degenerate nucleon system $\vert {\cal N}\ra$,
 for instance,
the quark condensate of the system 
can be given by,
\beq
 \langle {\cal N}\vert \bar{q}q\vert {\cal N}\rangle=
\frac{\partial\langle {\cal N}\vert {\cal H}_{QCD}
\vert {\cal N}\rangle}
{\partial m_q},
\eeq
where the expectation value of QCD Hamiltonian may be evaluated to 
\beq
\langle {\cal N}\vert {\cal H}_{QCD}\vert {\cal N}\rangle=
\varepsilon_{\rm vac}+\rho_{_B}[M_N+B(\rho_{_B})].
\eeq
Here, $\varepsilon_{\rm vac}$, $M_N$ 
and $B(\rho_{_B})$ denote the vacuum 
energy, the nucleon mass and the nuclear binding energy per particle,
respectively.
Thus one has
\beq
\langle \bar{q} q \rangle _{\rho_{_B}}
= \langle \bar{q} q \rangle_{_0}\cdot
\left[ 1 - {\rho_{_B} \over f_{\pi}^2 m_{\pi}^2 } \left( \Sigma_{\pi N}
 + \hat{m} {d \over d\hat{m}}  B(\rho_{_B}) 
\right)\right] ,
\eeq
where
$\Sigma _{\pi N}=(m_u+m_d)/2\cdot\la N\vert \bar{u}u+\bar {d} d\vert 
N\ra$ denotes the  $\pi$-N sigma term with $\hat{m} =(m_u + m_d)/2$;
the semi-empirical value of $\Sigma _{\pi N}$ is known to 
be $(40 - 50)$ MeV\cite{JK,bkm97}.
Notice that the correction term with finite $\rho_{_B}$ is positive
and  gives a reduction of almost 35 \% 
of $\langle \bar{q} q \rangle_{_0} $
already at the normal nuclear matter 
density $\rho_0 = 0.17 $fm$^{-3}$. 

If one believes in the above estimate\footnote{
It is said that one of mistakes which theoretical physicists
often do is {\em not} believing in their results with enough
strength and not pursuing
the consequences of them.}, one can 
conclude that
 the central region of heavy nuclei could be dense enough to  cause 
a partial restoration of chiral symmetry\cite{ptp95}, which may
induce  some characteristic phenomena 
of the chiral restoration in nuclear
medium\cite{hk94,BR}.

One may recall that
if a phase transition is of second order or weak first order,
there may exist specific  collective excitations called 
 {\em soft modes}\cite{soft}.
They are
the quantum fluctuations of the order parameter.
In the case of chiral transition, 
there are two kinds of fluctuations,
those of the phase and the modulus of the chiral condensate. 
The former is the Nambu-Goldstone boson, i.e., the pion, while 
the latter the $\sigma$  with the quantum numbers $I=0$ and 
$J^{PC}=0^{++}$\cite{ptp85}.

\setcounter{equation}{0}
\renewcommand{\theequation}{\arabic{section}.\arabic{equation}}

\section{The significance of the $\sigma$ meson
in low-energy hadron QCD}

 The  recent cautious phase shift
 analyses of the \pipi scattering have come to 
claim a  pole identified with
 the $\sigma$ in the $s$ channel together with the $\rho$ meson
pole in the $t$ channel\cite{pipiyitp,pipi,CLOSETORN}.
The $\sigma$ pole has  a real part 
Re\, $m_{\sigma}= 500$-600 MeV and the imaginary 
part Im\, $m_{\sigma}\simeq {\rm Re}\, m_{\sigma}$\cite{CLOSETORN}.
More recently, it has been  also found that
 the $\sigma$ pole gives a significant contribution to
 the decay processes of heavy particles involving a charm and 
$\tau$ leptons\cite{E791,bes,cleo}.
A  summary of the locations of the $\sigma$ pole in 
the complex energy plane may be found in \cite{XZ01}.

The elusiveness of the $\sigma$ meson comes from the fact that 
it strongly couples 
to two pions to acquire  a large width 
$\Gamma \sim m_{\sigma}$.
Although  the chiral perturbation theory
\cite{chipert} have made a great achievement 
for establishing the essential role of chiral symmetry
in describing (very) low-energy hadron phenomena,
 it is beyond the scope of the theory 
to  describe resonances\cite{OOR}.

The  important points in establishing the 
$\sigma$ pole consistently with  chiral symmetry is to incorporate
analyticity, unitarity {\em and} crossing symmetry\footnote{
For instance,
the phase shift in the $I=J=0$ channel could be well 
reproduced with a unitarized scattering amplitude 
lacking the $\sigma$ pole but including the
$\rho$ meson pole in the $t$-channel\cite{juerich,isgur}.}:
 Igi and Hikasa\cite{igi} constructed an
 invariant amplitude for the 
\pipi scattering using the $N/D$ method\cite{ND} 
so that it satisfies the chiral symmetry low energy 
theorem, analyticity, unitarity and especially 
(approximate) {\em crossing symmetry}.
They calculated  two cases with and without the scalar 
pole degenerated with the $\rho$ meson, the existence of which
was taken for granted.
What they found is that the $\rho$ only scenario can account only 
about a half of the observed phase shift, while the degenerate
$\rho$-$\sigma$ scenario gives an
excellent agreement with the
data.

In fact, the $\sigma$ meson  has been an enigma
in hadron physics\cite{CLOSETORN}:
It may be noticed, however, that the $\sigma$ meson 
as the quantum
fluctuation of the chiral order parameter
must be a {\em collective} state composed of many $q$-$\bar{q}$ states
 as the pion is\cite{nambu,ptp85}; notice that the pion can
 not be understood
within the conventional constituent quark model.

The possible existence of the $\sigma$ meson implies an
accumulation of the strength in the $I=J=0$ channel, so
is relevant to 
some observables\cite{hk94,ptp95};
(1) ~ $\Delta I=1/2$ rule in the kaon decay\cite{morozumi},
(2)~ the intermediate-range attraction in nuclear force
\cite{sawada,tamagaki},
 (3) ~  $\pi$-N sigma term\cite{kuni90} and so on.

\setcounter{equation}{0}
\renewcommand{\theequation}{\arabic{section}.\arabic{equation}}
\section{Partial chiral restoration and the 
$\sigma$ meson in  hadronic matter}

As was  first shown in\cite{ptp85}, 
if the $\sigma$ is really associated
with the fluctuation of the chiral order parameter,
one can  expect that the $\sigma$ pole moves toward
the origin in the complex energy plane
in the chiral limit and the $\sigma$ may become a sharp
resonance as chiral symmetry is restored 
 at high temperature and/or density; the $\sigma$ can be
a {\em soft mode}\cite{soft} of the chiral restoration; see also
\cite{bernard87}.

Some years ago\cite{tit,ptp95}, 
 several nuclear experiments including
 one using electro-magnetic probes 
were proposed to create the scalar mode in nuclei, thereby 
obtain a clearer evidence of  the  existence of 
the $\sigma$ meson and also examine the possible 
restoration of chiral  symmetry in  nuclear
 medium. It was also mentioned that
to avoid the huge amount of two pions from the $\rho$ meson, 
detecting  neutral pions through four $\gamma$'s may be convenient.

One must, however, notice that 
 a hadron put in a heavy nucleus may 
 dissociate into complicated
excitations to loose its identity.
Moreover, as remarked above, it is still uncertain whether
the $\sigma$ pole 
really corresponds to the pre-existing quantum fluctuation of the 
chiral order parameter or only a $\pi$-$\pi$ molecule generated
dynamically.
Then the most proper quantity to observe is the response function 
or spectral function in the channel with the same quantum number 
as the hadron has. 

Hatsuda, Shimizu and the present author (HKS) \cite{HKS}
showed that the spectral enhancement
near the $2m_{\pi}$ threshold takes place 
in association with  partial restoration of chiral symmetry  
at finite baryon density.
The calculation is a simple extension of the 
finite $T$ case\cite{CH98}.
The following should be, however, noticed;
since  HKS is based on a a perturbation theory
for treating
 the effects of the meson-loops as well as
 the baryon density,
 their loop-expansion  
  should be valid only at relatively low
 densities.
There are some attempts for developing the
chiral perturbation theory i.e., the non-linear realization
of chiral symmetry, for a finite density system
\cite{TW,mow,wirzba03}.

\setcounter{equation}{0}
\renewcommand{\theequation}{\arabic{section}.\arabic{equation}}
\section{Role of the wave function renormalization;
chiral restoration in nonlinear realization}

In the linear representation of chiral symmetry as given 
by the linear sigma model, it is rather apparant that
the possible chiral restoration affects 
 the dynamics in the medium because
 the $\sigma$ degree of freedom is 
explicit from the outset.
How can the possible chiral restoration be implemented in the 
non-linear realization where the $\sigma$ degree of freedom is
absent? 
Jido et al\cite{jhk}
showed that the nonlinear realization of the chiral 
symmetry can also
give rise to a near 2$m_{\pi}$ enhancement of the
spectral function in nuclear medium.
The enhancement of the cross section is found due to 
the wave function renormalization of the pion in nuclear medium
which causes the decrease of the pion decay constant 
$f_{\pi}^{\ast}(\rho)$
in the medium.
They identified that the wave function renormalization and hence 
the  decrease of  $f^{\ast}_{\pi}(\rho)$ are 
 owing to the following  new vertex:
\beq
\label{new-vertex}
{\cal L}_{\rm new} = - {3g \over 2 \lambda f_{\pi}} \
\bar{N}N {\rm Tr} [\partial U \partial U^{\dagger}].
\eeq

Here a couple of remarks are in order:
(i)~ The vertex eq.(\ref{new-vertex})
 has been known to be one of the next-to-leading order terms
 in the non-linear chiral Lagrangian in the
  heavy-baryon formalism  \cite{GSS}.
(ii)~The essential role of the  wave-function renormalization of 
the pion field on  the partial restoration of chiral symmetry 
 is  also shown  in the chiral perturbation theory in a more systematic
way\cite{TW,mow} and has been also 
  revealed  in accounting for the anomalous repulsion seen in the
deeply bound pionic nuclei\cite{kkw}. 

\section{The behavior of the $\sigma$ pole in the complex
energy plane}

 We  emphasize here 
 the relation of the near-threshold enhancement and the
 softening of the  fluctuating mode in the $\sigma$ channel in
 nuclear matter. We shall show that
such a fluctuating mode can be characterized by a
 complex pole of the unitarized scattering amplitude $T(s)$.

The  $T$ matrix resummed in the inverse amplitude method\cite{IAM} 
in the non-linear realization reads($ s=p^2$)
\beq
T(s)=T_2/(1+T_2g(s)),
\eeq
where $T_2=s/f^{\ast 2}_{\pi}$ and $g(s)$ is the pion loop integral;
\beq
g(s)=i\int\frac{d^4 q}{(2\pi)^4}\frac{1}{q_{+}^2-m_{\pi}^2+i\eps}
\frac{1}{q_{-}^2-m_{\pi}^2+i\eps},
\quad
q_{\pm}=q\pm p/2.
\eeq
Notice that  $g(s)$ satisfies the dispersion relation
\beq
g(s)=\frac{1}{\pi}\int \frac{{\rm Im}g(s')ds'}{s'-s-i\eps}.
\eeq
Here one can  easily find that
${\rm Im}g(s)=$
$-(1/ 16\pi) \sqrt{1-4m_{\pi}^2/s} \cdot \theta(s-4m_{\pi}^2)$.

The dispersion integral which is logarithumically divergent is
evaluated with  a simple cutoff.
 The analytic continuation of
the function $g(z)$ to the 2nd Riemann sheet (Im $z < 0$)
is given by
\beq
g(z)=
\frac{1}{\pi}\int_{\rm C'} \frac{{\rm Im}g(s')ds'}{s'-z},
\eeq
where C'$=$ C+ (a circle around $z$) with 
C being a straight line on $ 0 \leq z \leq \Lambda^2$.
The integral along the circle  gives an extra
term, $2\pi i$.
We remark that 
with this definition, $g(z)$ is continuous when $z$ crosses the 
 real axis from the upper  to the lower plane.

In the chiral limit,  one obtains
$g(z) = -\frac{1}{16\pi^2}\{{\rm Log}(1- \Lambda^2/z)+2\pi i\}$, 
(${\rm Im}z<0$).
Here ${\rm Log}z$ denotes the principal value of the logarithm, which
 has a cut along the real negative axis and 
$-\pi<{\rm Arg}[{\rm Log}z]<\pi$. 
The pole  is given as the solution to the following dispersion 
equation
$T_2^{-1}(z)=-g(z)$.
When $z$ is located in the 2nd sheet, the equation has the form
of
\beq
\label{disp_second}
16\pi^2f^{\ast 2}_{\pi}=z\{{\rm Log}(-\Lambda^2/z)+2\pi i\}.
\eeq
Let $z/\Lambda^2=a-ib$ \, ($a, b>0$),
and write $b/a=\tan \theta$, then
Im$[{\rm Log}(-\Lambda^2/z)]=-(\pi-\theta)$.
Thus eq.(\ref{disp_second}) is finally reduced to
$16\pi^2f^{\ast 2}_{\pi}=z\{-\log(\vert z\vert/\Lambda ^2)
+i\theta +i\pi \}$.
When 
$a, b \ll 1$, or $\vert z\vert/\Lambda ^2 \ll 1$ and
$\theta \ll 1$, we have
$16\pi^2f^{\ast 2}_{\pi}\simeq i\pi z$. This will give a check of
the numerical calculation.

The numerical solution to
eq.(\ref{disp_second}) is shown in Fig.1:\\
(1)~ A pole exists in the lower half plane in the 
 complex $s$ plane. (2)~ The pole
 moves toward the origin along the solid line
 as $f^{\ast}_{\pi}$=$\langle \sigma \rangle$ is decreased.
Thus one sees that 
 the mass and the width of the soft mode decreases
as the chiral symmetry is restored.

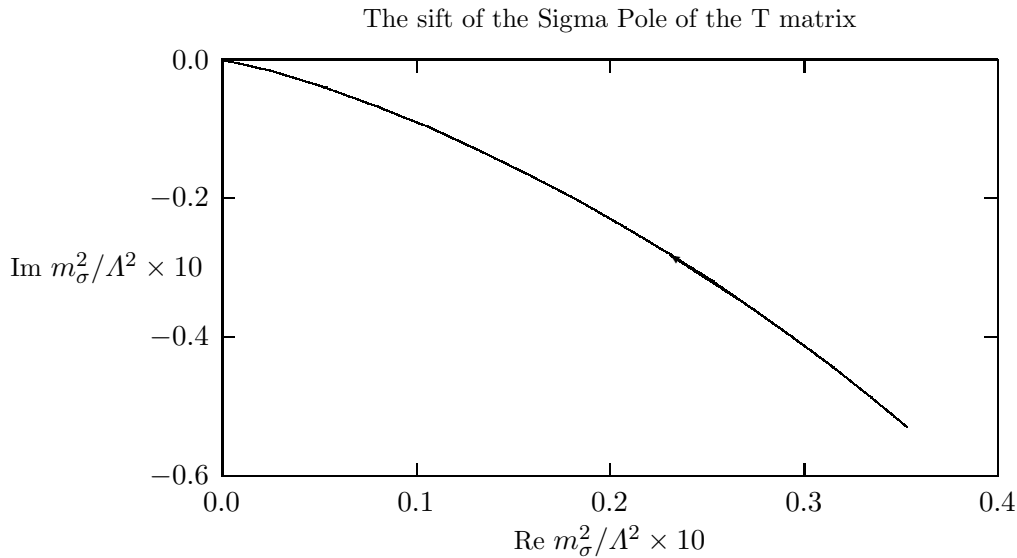
\begin{figure}
\begin{center}
\setlength{\unitlength}{0.240900pt}
\ifx\plotpoint\undefined\newsavebox{\plotpoint}\fi
\sbox{\plotpoint}{\rule[-0.200pt]{0.400pt}{0.400pt}}%
\begin{picture}(1500,900)(0,0)
\font\gnuplot=cmr10 at 10pt
\gnuplot
\sbox{\plotpoint}{\rule[-0.200pt]{0.400pt}{0.400pt}}%
\put(221.0,123.0){\rule[-0.200pt]{4.818pt}{0.400pt}}
\put(201,123){\makebox(0,0)[r]{$-0.6$}}
\put(1419.0,123.0){\rule[-0.200pt]{4.818pt}{0.400pt}}
\put(221.0,341.0){\rule[-0.200pt]{4.818pt}{0.400pt}}
\put(201,341){\makebox(0,0)[r]{$-0.4$}}
\put(1419.0,341.0){\rule[-0.200pt]{4.818pt}{0.400pt}}
\put(221.0,559.0){\rule[-0.200pt]{4.818pt}{0.400pt}}
\put(201,559){\makebox(0,0)[r]{$-0.2$}}
\put(1419.0,559.0){\rule[-0.200pt]{4.818pt}{0.400pt}}
\put(221.0,777.0){\rule[-0.200pt]{4.818pt}{0.400pt}}
\put(201,777){\makebox(0,0)[r]{$0.0$}}
\put(1419.0,777.0){\rule[-0.200pt]{4.818pt}{0.400pt}}
\put(221.0,123.0){\rule[-0.200pt]{0.400pt}{4.818pt}}
\put(221,82){\makebox(0,0){$0.0$}}
\put(221.0,757.0){\rule[-0.200pt]{0.400pt}{4.818pt}}
\put(526.0,123.0){\rule[-0.200pt]{0.400pt}{4.818pt}}
\put(526,82){\makebox(0,0){$0.1$}}
\put(526.0,757.0){\rule[-0.200pt]{0.400pt}{4.818pt}}
\put(830.0,123.0){\rule[-0.200pt]{0.400pt}{4.818pt}}
\put(830,82){\makebox(0,0){$0.2$}}
\put(830.0,757.0){\rule[-0.200pt]{0.400pt}{4.818pt}}
\put(1135.0,123.0){\rule[-0.200pt]{0.400pt}{4.818pt}}
\put(1135,82){\makebox(0,0){$0.3$}}
\put(1135.0,757.0){\rule[-0.200pt]{0.400pt}{4.818pt}}
\put(1439.0,123.0){\rule[-0.200pt]{0.400pt}{4.818pt}}
\put(1439,82){\makebox(0,0){$0.4$}}
\put(1439.0,757.0){\rule[-0.200pt]{0.400pt}{4.818pt}}
\put(221.0,123.0){\rule[-0.200pt]{293.416pt}{0.400pt}}
\put(1439.0,123.0){\rule[-0.200pt]{0.400pt}{157.549pt}}
\put(221.0,777.0){\rule[-0.200pt]{293.416pt}{0.400pt}}
\put(40,450){\makebox(0,0){Im  ${\small m_{\sigma}^2/\Lambda^2\times 10}$}}
\put(830,21){\makebox(0,0){Re $m_{\sigma}^2/\Lambda^2\times 10$}}
\put(830,839){\makebox(0,0){The sift of the Sigma Pole of the $$T$$ matrix}}
\put(221.0,123.0){\rule[-0.200pt]{0.400pt}{157.549pt}}
\multiput(1055.03,384.58)(-0.771,0.499){171}{\rule{0.716pt}{0.120pt}}
\multiput(1056.51,383.17)(-132.514,87.000){2}{\rule{0.358pt}{0.400pt}}
\put(924,471){\vector(-3,2){0}}
\put(221,777){\usebox{\plotpoint}}
\multiput(221.00,775.92)(2.228,-0.495){33}{\rule{1.856pt}{0.119pt}}
\multiput(221.00,776.17)(75.149,-18.000){2}{\rule{0.928pt}{0.400pt}}
\multiput(300.00,757.92)(1.630,-0.497){49}{\rule{1.392pt}{0.120pt}}
\multiput(300.00,758.17)(81.110,-26.000){2}{\rule{0.696pt}{0.400pt}}
\multiput(384.00,731.92)(1.364,-0.497){59}{\rule{1.184pt}{0.120pt}}
\multiput(384.00,732.17)(81.543,-31.000){2}{\rule{0.592pt}{0.400pt}}
\multiput(468.00,700.92)(1.249,-0.497){63}{\rule{1.094pt}{0.120pt}}
\multiput(468.00,701.17)(79.729,-33.000){2}{\rule{0.547pt}{0.400pt}}
\multiput(550.00,667.92)(1.086,-0.498){71}{\rule{0.965pt}{0.120pt}}
\multiput(550.00,668.17)(77.997,-37.000){2}{\rule{0.482pt}{0.400pt}}
\multiput(630.00,630.92)(1.017,-0.498){73}{\rule{0.911pt}{0.120pt}}
\multiput(630.00,631.17)(75.110,-38.000){2}{\rule{0.455pt}{0.400pt}}
\multiput(707.00,592.92)(0.965,-0.498){75}{\rule{0.869pt}{0.120pt}}
\multiput(707.00,593.17)(73.196,-39.000){2}{\rule{0.435pt}{0.400pt}}
\multiput(782.00,553.92)(0.871,-0.498){81}{\rule{0.795pt}{0.120pt}}
\multiput(782.00,554.17)(71.349,-42.000){2}{\rule{0.398pt}{0.400pt}}
\multiput(855.00,511.92)(0.835,-0.498){81}{\rule{0.767pt}{0.120pt}}
\multiput(855.00,512.17)(68.409,-42.000){2}{\rule{0.383pt}{0.400pt}}
\multiput(925.00,469.92)(0.792,-0.498){83}{\rule{0.733pt}{0.120pt}}
\multiput(925.00,470.17)(66.480,-43.000){2}{\rule{0.366pt}{0.400pt}}
\multiput(993.00,426.92)(0.723,-0.498){87}{\rule{0.678pt}{0.120pt}}
\multiput(993.00,427.17)(63.593,-45.000){2}{\rule{0.339pt}{0.400pt}}
\multiput(1058.00,381.92)(0.701,-0.498){87}{\rule{0.660pt}{0.120pt}}
\multiput(1058.00,382.17)(61.630,-45.000){2}{\rule{0.330pt}{0.400pt}}
\multiput(1121.00,336.92)(0.678,-0.498){87}{\rule{0.642pt}{0.120pt}}
\multiput(1121.00,337.17)(59.667,-45.000){2}{\rule{0.321pt}{0.400pt}}
\multiput(1182.00,291.92)(0.628,-0.498){91}{\rule{0.602pt}{0.120pt}}
\multiput(1182.00,292.17)(57.750,-47.000){2}{\rule{0.301pt}{0.400pt}}
\multiput(1241.00,244.92)(0.606,-0.498){91}{\rule{0.585pt}{0.120pt}}
\multiput(1241.00,245.17)(55.786,-47.000){2}{\rule{0.293pt}{0.400pt}}
\end{picture}
\end{center}
\caption{The movement of the $\sigma$ meson 
pole (in the chiral limit) as $f_{\pi}^{\ast}$
 decreases toward $0$.\cite{paris} }
\label{fig1}
\end{figure}

\section{Discussions}

\subsection{Experimental results on the spectral function
in the $\sigma$ channel}

There are a few relevant experiments which might show the 
softening of the spectral function in the $I=J=0$ channel.
\begin{enumerate}
\item
CHAOS collaboration  \cite{chaos}
 observed that
the   yield for  $M^A_{\pi^{+}\pi^{-}}$ 
 near the 2$m_{\pi}$ threshold 
 increases dramatically with increasing $A$.
They
identified that the $\pi^{+}\pi^{-}$ pairs in this range of
 $M^A_{\pi^{+}\pi^{-}}$ is in the $I=J=0$ state.
Although 
 this experiment was motivated to explore the $\pi$-$\pi$
correlations in nuclear medium\cite{motivation},
the near $2m_{\pi}$-threshold enhancement might be
attributed to a partial restoration of chiral symmetry in heavy
nuclei. Here it should be remarked, however,
that there are some attempts to explain the CHAOS data solely
by the many-body effects without
recourse to a possible vacuum change\cite{wambach}.
\item
The experiment to explore the spectral function in the same
channel in heavy nuclei were also performed by Crystal Ball
group\cite{CB}. The CHAOS group claims 
\cite{chaos} that there is no essential
difference between the two experiments, althou otherwise had been 
spelled out in \cite{CB}.
\item
A similar but more clear experimental result
 which shows a softening of the spectral
function in the $\sigma$ channel 
in the nuclear medium  has been also obtained by 
TAPS group\cite{taps}.
\end{enumerate}

\subsection{Deeply bound nuclei and chiral restoration}

The deeply bound pionic atom has proved 
to be a good probe of the properties of the hadronic interaction
 deep inside of heavy nuclei.  There is 
a suggestion \cite{WEISE,itahashi,kkw} that
the anomalously repulsive  energy shift
 of the pionic atoms (pionic nuclei)
owing to the strong interaction could be attributed to the
 decrease of $f^{\ast}_{\pi}(\rho)$ in heavy nuclei.
It may also imply that the chiral symmetry is partially restored deep
inside of nuclei.
A remarkable point is
 that the decrease of $f^{\ast}_{\pi}(\rho)$
is owing to  the wave-function
renormalization of the pion field in nuclei\cite{kkw},
as is for the in-medium $\pi\pi $ interaction in the
$I=J=0$ channel.

\section{Summary and concluding remarks}

The present report may be summarized as follows.\\
(1)~
Partial restoration of chiral symmetry in hot and dense medium
as represented by the decreasing $f_{\pi}^{\ast}$ leads to 
a shift
of the $\sigma$ meson pole in the 2nd Riemann sheet
 even in the
 non-linear realization of chiral symmetry.
(2)~The  decrease of $f_{\pi}^{\ast}$ in the nuclear medium
is a direct consequence of the wave function renormalization
of the pion field in the medium, which also accounts for the 
anomalous repulsion seen in the deeply bound pionic nuclei.
(3)~Even a slight restoration of chiral symmetry in the 
hadronic matter
 leads to a peculiar 
softening of the spectral function
 in the $\sigma$ channel.
(4)~ Such an enhancement might have  been observed 
 in the reactions creating two pions in nuclei.

In passing, some remarks are in order:
(a) Possible evidences of the 
partial restoration of chiral symmetry in hot and/or dense
 matter
are also obatined in the vector channel\cite{ceres,ozawa}.
(b)~The $N/D$ method applied to hot and/or dense medium
suggests that the $\sigma$ and the $\rho$ mesons can
become a soft mode simultaneously for the chiral 
restoration\cite{yokokawa}.  
(c)~ It is found\cite{kuni91,hi02} that a singular behavior inherent
to the  chiral transition also shows up in 
the baryon number susceptibility at finite chemical potential
$\mu_B\not=0$ 
owing to the scalar-vector mixing at $\mu_B\not=0$.

\acknowledgements
I thank  T. Hatsuda,
  H. Shimizu, D. Jido and K. Yokokawa
 for the collaboration and  discussions. 
This work is supported by the Grants-in-Aids of the Japanese
Ministry of Education, Science and Culture (No. 14540263).

\end{document}